\documentclass[a4paper,11pt]{article}
\usepackage{pos}
\usepackage{bm}
\usepackage{amsmath}
\usepackage{amsfonts}
\usepackage{amsthm}
\usepackage{physics}
\usepackage{bm}
\usepackage{slashed}
\usepackage{xcolor}
\usepackage[many]{tcolorbox}
\usepackage{booktabs, subfig}
\usepackage{graphics,epsfig}
\usepackage{hyperref}
\usepackage{multicol, multirow}
\usepackage[font=small,labelfont=bf]{caption}

\newcommand{\aem}{\alpha_{\mathrm{em}}}

\makeatletter
\newcommand{\subalign}[1]{%
  \vcenter{%
    \Let@ \restore@math@cr \default@tag
    \baselineskip\fontdimen10 \scriptfont\tw@
    \advance\baselineskip\fontdimen12 \scriptfont\tw@
    \lineskip\thr@@\fontdimen8 \scriptfont\thr@@
    \lineskiplimit\lineskip
    \ialign{\hfil$\m@th\scriptstyle##$&$\m@th\scriptstyle{}##$\hfil\crcr
      #1\crcr
    }%
  }%
}
\makeatother

\title{Valence leading isospin breaking contributions to $a_{\mu}^{\mathrm{HVP-LO}}$}

\author[a]{Simone Bacchio}
\author*[b]{Antonio Evangelista}
\author[b]{Roberto Frezzotti}
\author[c,d]{Giuseppe Gagliardi}
\author[e]{Marco Garofalo}
\author[f]{Nikolaos Kalntis}
\author[f]{Simone Romiti}
\author[d]{Francesco Sanfilippo}
\author[b]{Nazario Tantalo}

\affiliation[a]{Computation-based Science and Technology Research Center, The Cyprus Institute, 20 Konstantinou Kavafi Street, 2121 Nicosia, Cyprus}
\affiliation[b]{Dipartimento di Fisica and INFN, Università di Roma “Tor Vergata", Via della Ricerca Scientifica 1, I-00133 Rome, Italy}
\affiliation[c]{Dipartimento di Matematica e Fisica, Università Roma Tre, Via della Vasca Navale 84, I-00146  Rome, Italy}
\affiliation[d]{Istituto Nazionale di Fisica Nucleare, Sezione di Roma Tre, Via della Vasca Navale 84, I-00146  Rome, Italy}
\affiliation[e]{HISKP (Theory), Rheinische Friedrich-Wilhelms-Universität Bonn, Nussallee 14-16, 53115 Bonn, Germany}
\affiliation[f]{Institute for Theoretical Physics, Albert Einstein Center for Fundamental Physics, University of Bern, CH-3012 Bern, Switzerland}

\onbehalf{for the Extended Twisted Mass collaboration}

\emailAdd{antonio.evangelista@roma2.infn.it}

\abstract{
{\centering
\includegraphics[width=0.18\textwidth]{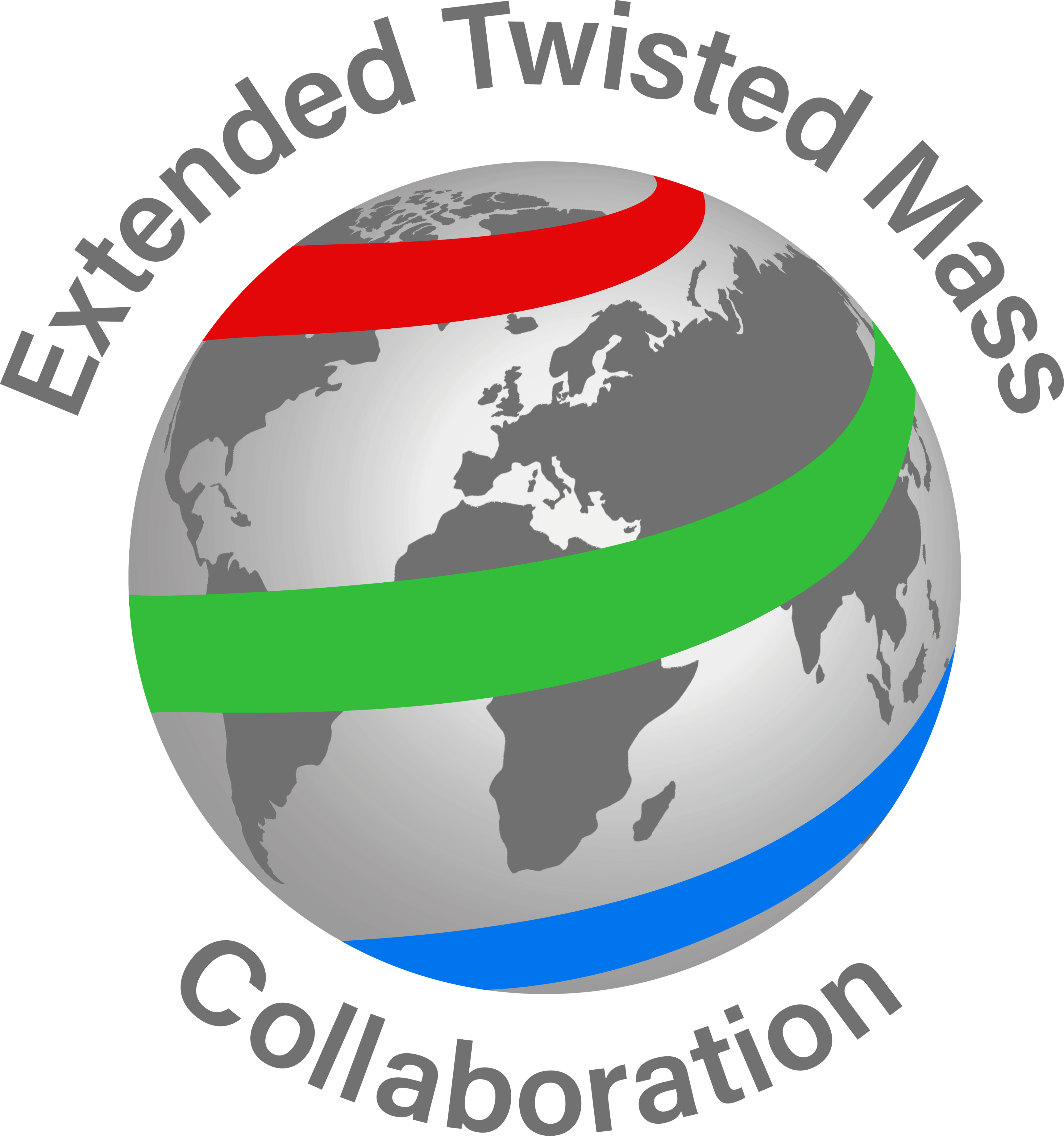} \\[5pt]
}
By employing the RM123 approach to QCD+QED, we computed the valence quark-connected isospin-breaking corrections to the light, strange and charm contributions at leading order in $\alpha_{\mathrm{em}}$ and $\left(\mu_d-\mu_u\right) / \Lambda_{\mathrm{QCD}}$.
Here we report the preliminary results on two different volumes ($L \sim 3.8$~fm and $L \sim 5.1$~fm) and a fixed lattice spacing (corresponding to $a_{\text {isoQCD }} \sim 0.07951(4)$~fm), obtained in the framework of the ongoing computation by ETMC of the leading-order hadronic vacuum polarization (HVP) contribution to the muon anomalous magnetic moment $a_\mu^{\mathrm{HVP}}$ in QCD+QED.
}

\FullConference{%
The 41st International Symposium on Lattice Field Theory (Lattice2024)\\
27 July - 03 August, 2024\\
University of Liverpool, Liverpool, United Kingdom
}

\begin{document}
\maketitle
\section{Introduction}

In the last few years the lattice QCD community has made impressive progress in increasing the precision of the Hadronic Vacuum Polarization (HVP) contribution to the anomalous magnetic moment of the muon $a_{\mu}^{\mathrm{HVP}}$, reaching the few permille accuracy. The Leading Isospin Breaking (LIB) corrections, coming from both QED and strong isospin breaking (SIB) effects, are of first order in the small parameters $\alpha_{\mathrm{em}}$ and $(\mu_d -\mu_u)/\Lambda_{\mathrm{QCD}}$ but relevant at the subpercent accuracy. They have been fully computed by the BMW collaboration \cite{Borsanyi:2020mff} and partly by other groups \cite{Djukanovic:2024cmq, FermilabLattice:2024yho}.

In this work, we present the status of LIB effects computation by ETM Collaboration. By using the RM123 approach \cite{deDivitiis:2013xla}, we focused on the electro-quenched (EQ) light, strange and charm connected contributions, while work on the EQ disconnected and the electro-unquenched contributions is in progress. We computed EQ contributions on two ensembles with equal lattice spacing. $a\sim 0.08$ fm, and aspect ratio, $T/L=2$, but different linear size, i.e.\ $L \sim 3.8$~fm and $L \sim 5.1$~fm.
\section{Lattice setup}
\label{Sec:1}

We outline here the mixed action lattice setup and the RM123 expansion method we employ to evaluate correlation functions in QCD+QED to the first order in $\alpha_{em}$ and $\mu_d -\mu_u$, using twisted mass fermions and the ensembles produced by ETM in isospin symmetric QCD (isoQCD).
We define the QCD+QED lattice theory through the $U(1) \times SU(3)$ gauge-invariant and flavour diagonal action
\begin{equation}
\!\!\! S_{\mathrm{QCED}}[U, A, q_f, \bar{q}_f] \! = \!\! \sum_{\subalign{f= & u, d \\ & s, c}}\! \sum_{x} \bar{q}_{f}(x) \! \left\{ \gamma_\mu \Tilde{\nabla}_\mu\left[U, A\right] - i \gamma_5 \, r_f \left(W_{cl}\left[U, A\right] + m_{\mathrm{cr}}^f\right) + \mu_f \right\}\! q_{f}(x)\ ,
\label{Eq:dia_action}
\end{equation}
where $r_u=r_c=1$ and $r_d=r_s=-1$, and $m_{\mathrm{cr}}^f$ is the critical mass for flavour $f$. We set $m_{\mathrm{cr}}^{c}=m_{\mathrm{cr}}^{u}$ and $m_{\mathrm{cr}}^{s}=m_{\mathrm{cr}}^{d}$ 
since the shift of critical masses arises only from QED effects. 
Furthermore, it can be proved (following closely the arguments in~\cite{Frezzotti:2005gi}) that all physical observables of QCD+QED can be extracted without $\order{a^{2k+1}}$ lattice artifacts ($k$ integer positive) from correlators evaluated using the action $S_{\mathrm{QCED}}$. 

On the other hand, since the action $S_{\mathrm{QCED}}$ in general has a complex fermionic determinant and is not suitable for Markov chain based simulations, we employ a mixed action lattice setup that, in analogy to the discussion in App.~A of~\cite{ExtendedTwistedMassCollaborationETMC:2024xdf}, can be sketched as follows
\begin{equation}
S_{\mathrm{mix}} = S_{\mathrm{TM}   }\qty[\vec{g}^0] + 
                   S_{\mathrm{ghost}}\qty[\vec{g}^0] +
                   S_{\mathrm{QCED} }\qty[\vec{g}]\, ,
\end{equation}
where $S_{\mathrm{TM}}\qty[\vec{g}^0]$ denotes the TM isoQCD action for two pairs of maximally twisted Wilson fermions, with the $c$--$s$ pair sector being flavour non-diagonal due to its Wilson term;
$S_{\mathrm{ghost}   }\qty[\vec{g}^0]$ refers to isoQCD too, but has the same flavour diagonal form as $S_{\mathrm{QCED} }\qty[\vec{g}]$,
which however carries the QCD+QED parameters. Here the bare parameters for isoQCD  and QCD+QED are denoted~\footnote{As we work in the EQ approximation we assume $a=a^0$ (equal lattice spacing in the two theories) with a change in the bare strong coupling.} by  
\begin{equation}
\vec{g}^0 = \left\{\left(g_S^0\right)^2, 0, \mu_{\ell}^{0}, \mu_{\ell}^{0}, \mu_{s}^{0}, \mu_{c}^{0}; m_{\mathrm{cr}}^{0}, m_{\mathrm{cr}}^{0}\right\}\;, \qquad
\vec{g} = \left\{\left(g_S\right)^2, e^2, \mu_{u}, \mu_{d}, \mu_{s}, \mu_{c}; m_{\mathrm{cr}}^{u}, m_{\mathrm{cr}}^{d}\right\}\,,
\end{equation}
and $\order{a}$ improved results for all physical observables are expected based on the same arguments as mentioned above. The electro-unquenched LIB corrections to isoQCD can be obtained by evaluating the first--order expansion of the ratio of Dirac operator determinants from the actions
$S_{\mathrm{QCED} }\qty[\vec{g}]$ and $S_{\mathrm{ghost}}\qty[\vec{g}^0]$.

In this work, we applied the RM123 method to compute LIB corrections. Indeed, since the difference $\vec{g} - \vec{g}^0$ is small we can expand the QCD+QED theory around the isoQCD one, up to the first--order in $\aem$ and $\delta_{ud} \equiv (\mu_{u} - \mu_{d})\slash \Lambda_{\mathrm{QCD}}$. The expectation values in QCD + QED $\expval{\cdot}^{\vec{g}}$ is given by
\begin{equation}
\label{expansion}
\expval{\mathcal{O}\left[U,A; \vec{g}   \right]}^{\vec{g}} = 
\expval{\mathcal{O}\left[U;   \vec{g}_0 \right]}^{\vec{g}_0} 
+ \Delta \left\langle {\cal O}[U,A; \vec{g} \, ] \right\rangle ^{\vec{g}_0}  + 
\order{\aem^2, , \delta_{ud}^2, \aem \delta_{ud}}
\; .
\end{equation}
where the $\Delta$ operator is defined by
\begin{equation}
\Delta \expval{\mathcal{O}}^{\vec{g}_0} = 
\frac{1}{2}e^2  \eval{\pdv{\expval{\mathcal{O}}^{\vec{g}}}{e^2}}_{\vec{g} = \vec{g}_0}   
+ \sum_{\subalign{f= & u,d \\ & s,c}} (\mu_f-\mu_f^0) \eval{\pdv{\expval{\mathcal{O}}^{\vec{g}}}{m_f}}_{\vec{g} = \vec{g}_0}
+ \sum_{\subalign{f= & u,d \\ & s,c}} (m_{\mathrm{cr}}^{f} - m_{\mathrm{cr}}^0) \eval{\pdv{\expval{\mathcal{O}}^{\vec{g}}}{m_{\mathrm{cr}}^f}}_{\vec{g} = \vec{g}_0} \,.  
\label{Eq:expansion}
\end{equation}
\begin{table}[t]
\centering
\small
\begin{tabular}{ccccccccc}
\toprule
ensemble & $\beta$ & $V\slash a^4$ & $a^{\text{sim}}$ & $a\mu_{\ell}$ & $a\mu_{s}$ & $a \mu_{c}$ & $M_{\pi}$ (MeV) & $M_{\pi}L$ \\
\midrule
 B48 & 1.778 & $48^3 \cdot 96$  & 0.08 & 0.00072 & 0.01825  & 0.2377 & 141 & 2.7 \\
 B64 & 1.778 & $64^3 \cdot 128$ & 0.08 & 0.00072 & 0.01825  & 0.2377 & 140 & 3.6 \\
\midrule
\midrule
 \multirow{2}{*}{ensemble} & \multirow{2}{*}{$N_{\text{U}}$} & \multicolumn{7}{c}{$N_{\eta}$} \\
\cmidrule{3-9}
 & & $\mu_\ell$ & $3\mu_\ell$ & $5\mu_\ell$ & $7\mu_\ell$ & $9\mu_\ell$ & $\mu_s$ & $\mu_c$ \\
 \midrule
 B48 & 438 & 96 & 96 & 96 & 96 & 96 & 96 & 4 \\
 B64 & 417 & 64 & 64 & 20 &  8 &  4 & 64 & 4 \\
\bottomrule
\end{tabular}
\label{Tab:isoQCD_par}
\caption{We provide the parameters of the ETMC gauge ensembles used in this work, together with the number of configurations $N_{\text{U}}$ analysed and the number of stochastic sources $N_{\eta}$ employed to compute the quark propagators at each quark mass.}
\end{table}

In this work, we used two different strategies to compute the derivatives w.r.t $e^2$  w.r.t  $\mu_f$ and $m_{\mathrm{cr}}^{f}$.
Concerning derivatives w.r.t. $e^2$ we performed a finite difference approximation of the second derivatives, using values of $e^2=10^{-2}, 10^{-3}$ which are small enough to make $\order{e^4}$ in the differences fully negligible.
On the other hand, we implemented directly the scalar and pseudoscalar integrated density operator insertions corresponding, respectively, to the derivatives w.r.t. $\mu_f$ and $m_{\mathrm{cr}}^{f}$.
\section{Counterterms determinations}
\label{Sec:2}
In this section, we describe the determination of the counterterms needed to compute the corrections to all physical observables. We will first discuss the shift of critical masses and then the determination of physical quark mass counterterms.

To determine $m_{\mathrm{cr}}^{u} - m_{\mathrm{cr}}^{0}$ and $m_{\mathrm{cr}}^{d} - m_{\mathrm{cr}}^{0}$ in general we have to exploit two independent conditions, which here we choose by requiring restoration of parity invariance \cite{Frezzotti:2016lwv}.
\begin{align}
\Bar{C}(t)  & \equiv \; \partial_0  \sum_{\vec{x}}\left\langle \, V_0^1(x)P^2(0) \, \right\rangle^{\vec{g}}\big\slash  2\sum_{\vec{x}} \left\langle \, P^2(x)P^2(0) \, \right\rangle^{\vec{g}}= 0
\label{Eq:mcr_cond1}\\
C_{SP}(t)  & \equiv \sum_{\vec{x}}\left\langle \, S^1(x)P^1(0) \, \right\rangle^{\vec{g}}  = 0\;,
\label{Eq:mcr_cond2}
\end{align}
where $V_0^1(y) = \left( \Bar{\psi}_\ell\, \gamma^0\, \tau_1\,  \psi_\ell \right)(y)$, $P^{1,2}(y) = \left( \Bar{\psi}_\ell\, \gamma^5\, \tau_{1,2}\,  \psi_\ell \right)(y)$ and $S^{1}(y) = \left( \Bar{\psi}_\ell\, \bm{1} \, \tau_{1}\,  \psi_\ell \right)(y)$ with $\psi_{\ell} = (u, d)^{T}$.
We recall that in isoQCD the condition of Eq.~(\ref{Eq:mcr_cond1}) is used to determine $m_{\mathrm{cr}}^0$. Furthermore, as here we work in the EQ approximation, there is no contribution from sea quark effects and hence one has $\Delta m_{\mathrm{cr}}^{d} = \left(e_d\slash e_u\right)^2 \Delta m_{\mathrm{cr}}^{u}$ and can define $\Delta \Bar{m}_{cr} \equiv \left(e_u^2 + e_d^2\right)\left(\Delta m_{\mathrm{cr}}^{f}\slash 2e_f^2\right)$ ($e_f$ is the charge of the quark of flavour $f$).
We then extract the critical mass shifts $\Delta m_{cr}^{u, d}$ by fitting the lattice estimator $\Delta \Bar{m}_{cr}$ namely 
\begin{align}
\Delta \Bar{m}_{cr} (t) \equiv - \frac{e_u^2 + e_d^2}{2}\frac{e^2 \pdv{\Bar{C}(t) }{e^2}}{\left[e_u^2\pdv{\Bar{C}(t)}{m_{\mathrm{cr}}^{u}} + e_d^2\pdv{\Bar{C}(t)}{m_{\mathrm{cr}}^{d}}\right]}
\label{Eq:mcr}
\end{align}
to a constant ansatz in a time plateaux region that we conveniently choose as $1.0~\text{fm} \le t  \le 2.5~$~fm. 

\begin{figure}[t] 
\centering
\includegraphics[width=0.49\linewidth]{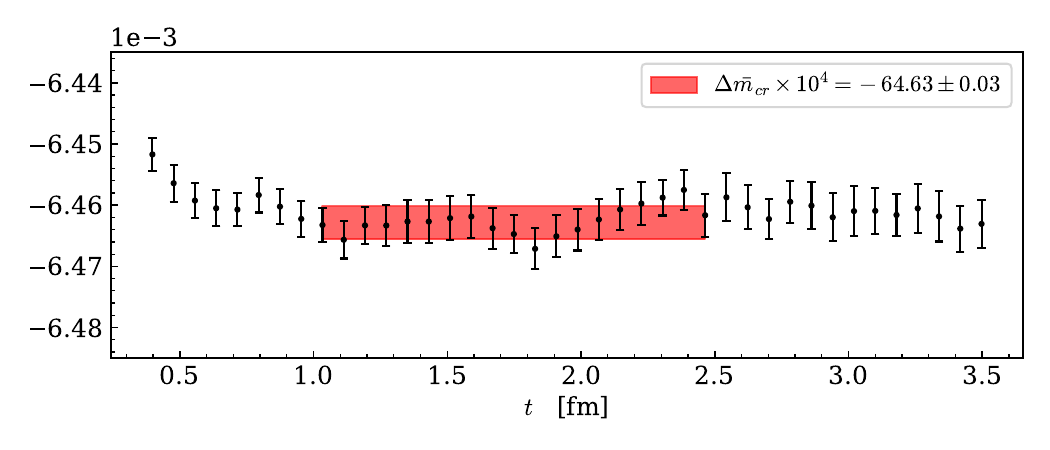}
\includegraphics[width=0.49\linewidth]{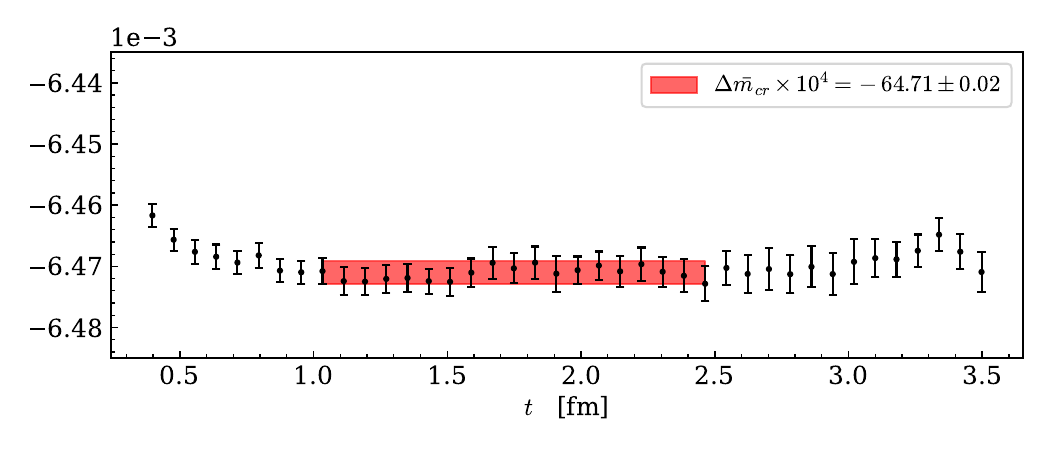}
\caption{Critical mass shift $\Delta \Bar{m}_{cr}$ determinations on B48 (left panel) and B64 (right panel).}
\label{Fig:mcr}
\end{figure}

By applying the $\Delta$ operator to the masses $M_P$ of the mesons $\pi^{+}$, $K^{+}$, $K^{0}$ and $D_s$ mesons, we determine the shift of the bare physical masses by solving the system for $\Delta \mu_f = \mu_f - \mu_f^0$, $f = u,d,s,c$,
\begin{align}
\sum_f \Delta \mu_f \pdv{M_P(\beta)}{\mu_f} = \left\{aM_P^{\mathrm{exp}} - \left[aM_P\right]^{\mathrm{iso}}\right\} - e^2\pdv{M_P(\beta)}{e^2} - \sum_f \Delta m_{\mathrm{cr}}^{f} \pdv{M_P(\beta)}{m_{cr, f}}\,,
\label{Eq:system}
\end{align}
where $a = a^{\mathrm{iso}}$ is the isoQCD lattice spacing. Furthermore, we use the QED Finite Size Effects (FSE) relation \cite{BMW:2014pzb}
\begin{equation}
M_P(L) - M_P = -\frac{(Qe)^2 c}{8\pi} \left[ \frac{1}{L} + \frac{2}{M_PL^2} + O(L^{-3}) \right] \quad \text{with} \quad c = 2.837297 \ldots
\end{equation}
to correct for FSE in the term $[aM_P^{\mathrm{exp}} - \left[aM_P\right]^{\mathrm{iso}}]$ of Eq.~(\ref{Eq:system}).
We define our isoQCD theory according to the Edinburgh/FLAG consensus
\begin{gather}
 f_{\pi} = 130.5~\text{MeV} \qquad M_{\pi} = 135.0~\text{MeV} \qquad M_{K} = 494.6~\text{MeV} \qquad M_{D_s} = 1967~\text{MeV} 
\end{gather}

Before presenting the results for these counterterms, some remarks are in order.
The system of Eq.~(\ref{Eq:system}) is understood to hold in the EQ approximation. Beyond this approximation, LIB effects change the lattice spacing too, and a further input, such as $M_{\Omega}$ or $M_{D^0}$, would be required.

The bare physical quark mass shifts $\Delta \Bar{\mu} = \left(\Delta \mu_d + \Delta \mu_u\right)\slash 2$, $\Delta \mu_{ud} = \left(\Delta \mu_d - \Delta \mu_u\right)\slash 2$, $\Delta \mu_s$, $\Delta \mu_c$, as well the EQ critical mass shift $\Delta \Bar{m}_{cr}$ (see Tab.~\ref{Tab:counterterms}) are evaluated on lattices $L^3 T$ of two different linear sizes, $L\sim3.8$~fm (B48) and $L\sim5.1$~fm (B64) respectively. They agree within about two standard deviations, although the errors quoted here are merely statistical and a careful estimate of systematic errors  is deferred to future work. 

\begin{table}[t]
\centering
\begin{tabular}{rcccccc}
\toprule
    & $a\Delta \Bar{m}_{cr} \times 10^4$ & $a\Delta \Bar{\mu} \times 10^4$ & $a\Delta \mu_{ud} \times 10^4$ & $a\Delta \mu_s \times 10^4$ & $a\Delta \mu_c \times 10^4$ \\
\midrule
B48    & $-64.63 \pm 0.03$ & $-0.201 \pm 0.002$ & $2.65 \pm 0.03$ & $-0.436 \pm 0.006$ & $-23.72 \pm 0.06$ \\
B64    & $-64.71 \pm 0.02$ & $-0.196 \pm 0.001$ & $2.56 \pm 0.02$ & $-0.446 \pm 0.004$ & $-23.63 \pm 0.04$ \\
\bottomrule
\end{tabular}
\caption{The counterterms $\Delta \mu_f$ obtained solving for in Eq.~\ref{Eq:system}. The results are shown on the two volumes, $L\sim3.8 $~fm (B48) and $L\sim5.1 $~fm (B64). 
}
\label{Tab:counterterms}
\end{table}
\section{The LIB corrections to \texorpdfstring{$a_{\mu}$}{amu}}
\label{Sec:3}

In this section, we describe how to compute the LIB corrections for the light, strange and charm HVP contributions to the anomalous muon's magnetic moment. 
As customary, we adopt the time-momentum representation \cite{Bernecker:2011gh},
\begin{equation}
a_\mu^{\mathrm{HVP}}=2 \alpha_{\mathrm{em}}^2 \int_0^{\infty} \dd{t}\, K\left(m_\mu t\right) \Hat{C}_{JJ}(t) \,,
\label{Eq:amu}
\end{equation}
where $K(m_{\mu}t)$ is the usual analytic kernel and $\Hat{C}_{JJ}(t)$ is the renormalized two-point vector-vector correlator that can be decomposed on the flavour basis 
\begin{gather}
\Hat{C}_{JJ}^f(t) = Z_{V,f}^2 C_{JJ}^f(t)  \qquad
C_{JJ}^f(t) = - \frac{1}{3}e_f^2\sum_{k=1}^{3}\expval{J_k^f(x)J_k^f(0)}^{\vec{g}}\,.
\end{gather}
Here $Z_{V, f}$ is the renormalization constant for the flavour $f$ needed to match the local vector current to the conserved one. The correction to the renormalized two-point vector-vector correlator reads
\begin{align}
\Delta \Hat{C}_{JJ}^f(t) & \equiv \Hat{C}_{JJ}^f(t) - Z_{V,f}^2 C^{f, \mathrm{iso}}_{JJ}(t) \nonumber \\
    & = Z_{V,f}^2 \Delta C_{JJ}^f(t) + 2 Z_{V}\Delta  Z_{V,f} C^{f, \mathrm{iso}}_{JJ}(t)
\end{align}
where $\Delta Z_{V, f}$ is the QED correction for the flavour $f$ to the isoQCD renormalization constant $Z_V$ and $\Delta C_{JJ}^f(t)$ is the correction of the bare vector-vector two-point correlator, which according to Eq.~(\ref{Eq:expansion}) reads
\begin{equation*}
\Delta C_{JJ}^f(t) = 
e_f^2\,\pdv{C_{JJ}^f}{e^2}\,(t) 
+
\Delta m_{cr}^f\,\pdv{C_{JJ}^f}{m_{cr}^f}\,(t)
+ 
\Delta \mu_f\, \pdv{C_{JJ}^f}{\mu_f}\,(t)
\end{equation*}

Before computing the correction to the light, strange and charm contribution to the HVP, we discuss the determination of the correction to the vector current renormalization constant. Following the discussion of Appendix B of \cite{ExtendedTwistedMass:2022jpw}, and noticing that the relation used to determine $Z_{V, f}$ normalization constant in isoQCD keeps holding in our QCD + QED mixed action setup for a $SU(2)$ doublet of valence quarks of equal charge and opposite Wilson parameter ($r_{f_2} = -r_{f_1}$), we can compute $\Delta Z_{V, f}$ as follows 
\begin{equation}
\Delta Z_{V,f} = \Delta\left\{2\mu_{f}\frac{C_{PP}(t)}{\partial_0 C_{A_0P}(t)}\right\}
\qquad
\Delta Z_{V} \equiv \frac{1}{2}\left(\frac{\Delta Z_{V, u}}{e_u^2} + \frac{\Delta Z_{V, d}}{e_d^2}\right)\,.
\end{equation} 

\begin{figure}[t]
\centering
\includegraphics[width=0.49\linewidth]{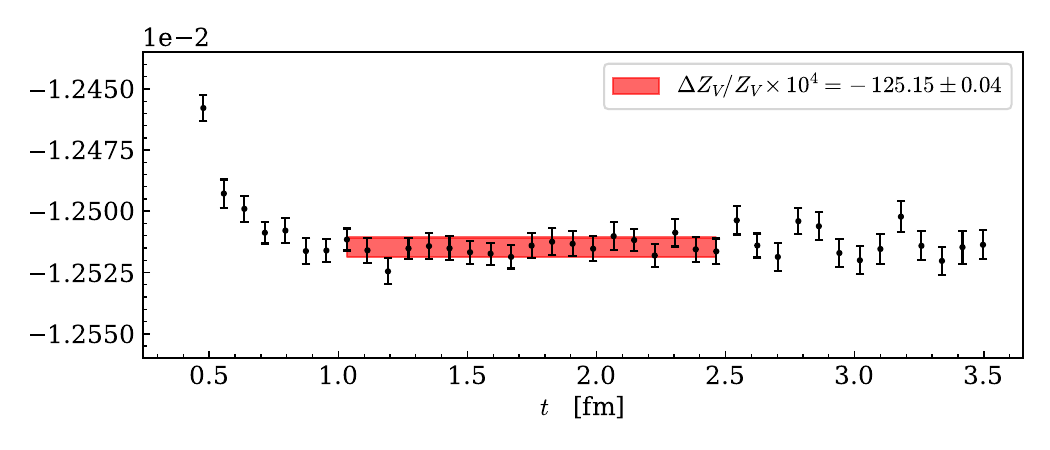}
\includegraphics[width=0.49\linewidth]{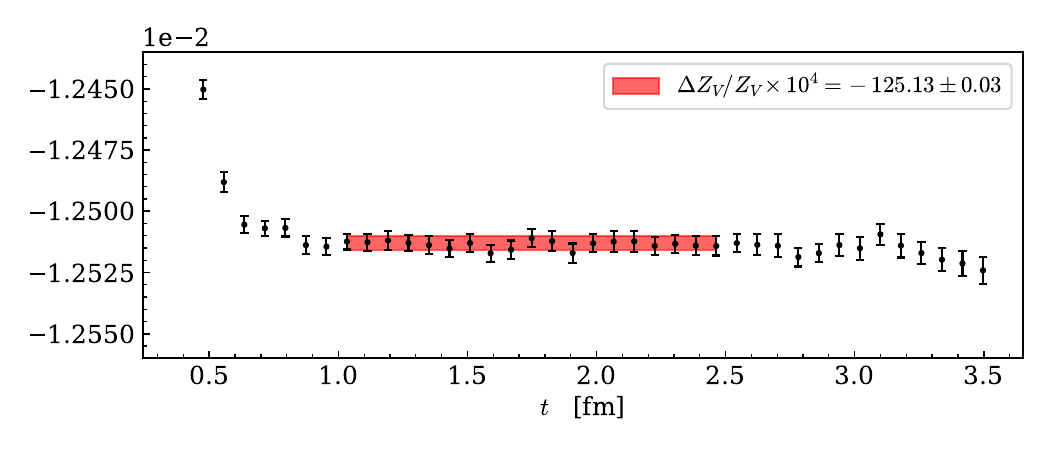}
\caption{The time behaviour of the estimator of $\Delta Z_{V} \slash Z_{V}$ and the constant fit results are shown for B48 (left panel) and B64 (right panel).}
\label{Fig:dZ_determ}
\end{figure}

In Fig.~\ref{Fig:dZ_determ} we show the time behaviour of the estimator of $\Delta Z_{V} \slash Z_{V}$. The results, obtained by fitting with a constant ansatz in the plateaux region between $t\sim1.0$~fm and $t\sim2.5$~fm, are 
\begin{equation}
\Delta Z_{V} \slash Z_{V} \times 10^4 = -125.14 \pm 0.04\,, \qquad 
\Delta Z_{V} \slash Z_{V} \times 10^4 = -125.13 \pm 0.03\,
\end{equation}
for B48 and B64 respectively.

\subsection{The LIB correntions to \texorpdfstring{$a_{\mu}^{\mathrm{HVP}}\left(\ell\right)$, $a_{\mu}^{\mathrm{HVP}}\left(s\right)$ and $a_{\mu}^{\mathrm{HVP}}\left(c\right)$ quark--connected contributions}{light contribution to the HVP}}

For each flavour the correction to $a_{\mu}^{\mathrm{HVP}}(f)$ is estimated from the limit for $t_{\text{cut}} \mapsto \infty$ of
\begin{equation}
\Delta a_{\mu}^{\mathrm{HVP}}(f; t_{\mathrm{cut}}) = 
2 \alpha_{\mathrm{em}}^2 \sum_{n=0}^{n_{\mathrm{cut}}} n^2 K\left(m_\mu an\right) \Delta \Hat{C}_{JJ}^{f}(an)\,,
\label{Eq:Damu}
\end{equation}
where $n = t\slash a$ is the Euclidean time in lattice units. 

\begin{figure}[t] 
\centering
\includegraphics[width=0.49\textwidth]{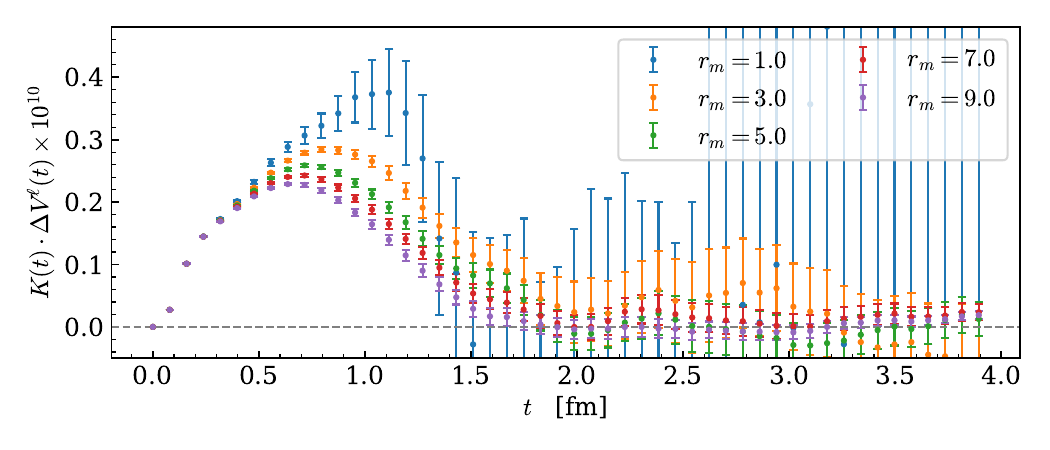}
\includegraphics[width=0.49\textwidth]{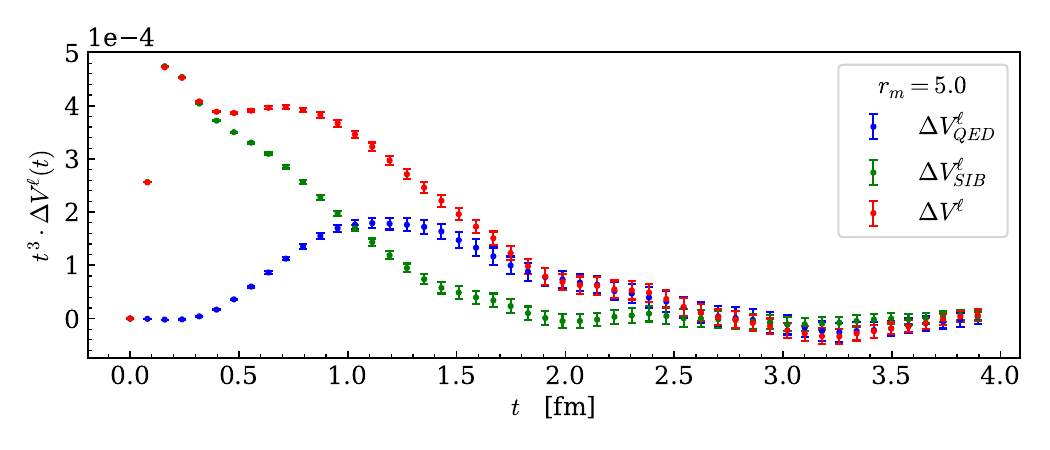}
\includegraphics[width=0.49\textwidth]{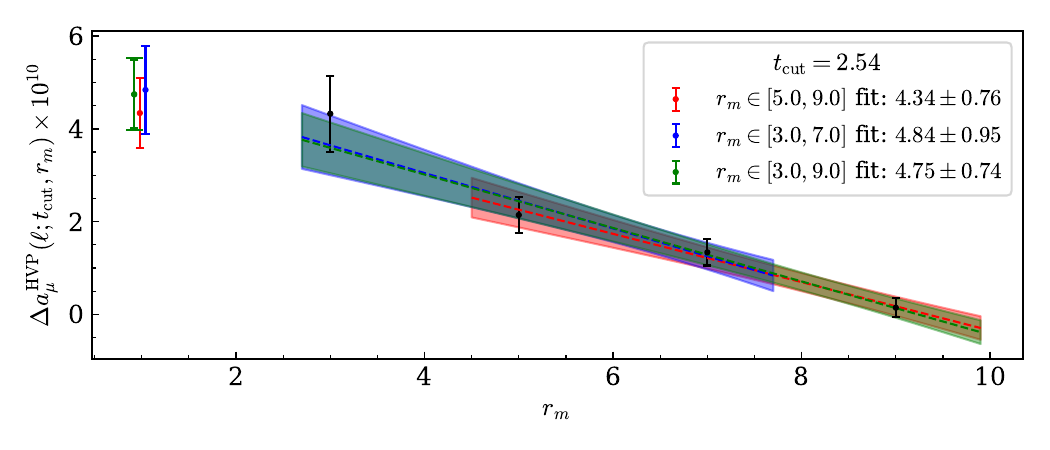}
\includegraphics[width=0.49\linewidth]{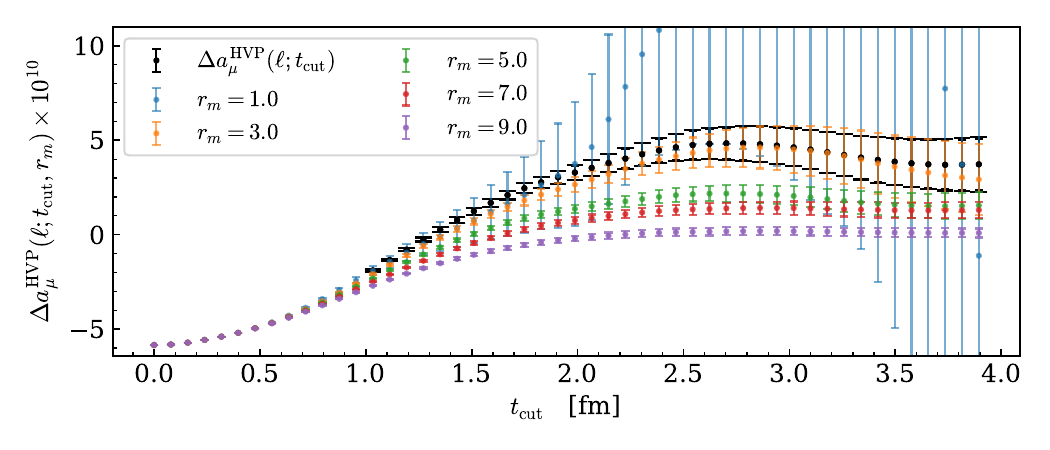}
\caption{In the case of ensemble B64, we show the integrand of Eq.~(\ref{Eq:Damu}) for $r_{m}=3, 5, 7, 9$ and the SIB and QED correction to $V^{u}(t) + V^{d}(t)$ with $r_{m}=5$ (left and right top panel, respectively). The left and right bottom panels show the chiral extrapolation and the time dependence of the simulated and extrapolated $\Delta a_{\mu}^{\mathrm{HVP}}(\ell; t_{\mathrm{cut}})$.}
\label{Fig:light}
\end{figure}

While for the charm and strange contributions, vector-vector two-point correlators are very precise, for the up and down contributions these are very noisy at large Euclidean time. To reduce the errors we performed the calculation of $a_{\mu}^{\mathrm{HVP}}\left(\ell\right)$ for four different light quark masses, $m_{\ell} = r_{m} \mu_{\ell}$ with $r_{m} = 3, 5, 7, 9$, exploiting a chiral extrapolation to obtain the target isoQCD light point ($r_m=1$) (this procedure has been proposed and used in \cite{Borsanyi:2020mff}, see Fig.~\ref{Fig:light}). For each $t_{\mathrm{cut}}$, the chiral extrapolation has been performed using the ansatz
\begin{equation}
\Delta a_{\mu}^{\mathrm{HVP}}(f; t_{\mathrm{cut}}, r_m) = \Delta a_{\mu}^{\mathrm{HVP}}(f; t_{\mathrm{cut}}) + c_1\cdot r_{m}
\end{equation}
using all the data--points, only those corresponding to $r_{m}=(3, 5, 7)$ and only those corresponding to $r_{m} = (5, 7, 9)$ (see left bottom panel of Fig.~\ref{Fig:light}). The first kind of fit is used to quote the central values, while we estimated the systematic errors on the extrapolation from the spread between the second and third kind of fit.
Finally, the $t_{\text{cut}} \mapsto \infty$ is performed fitting to a constant in the region between $t_{\mathrm{cut}}\sim 2.5$~fm and $t_{\mathrm{cut}}\sim 3.5$~fm. 

\begin{figure}[t] 
\centering
\includegraphics[width=0.49\linewidth]{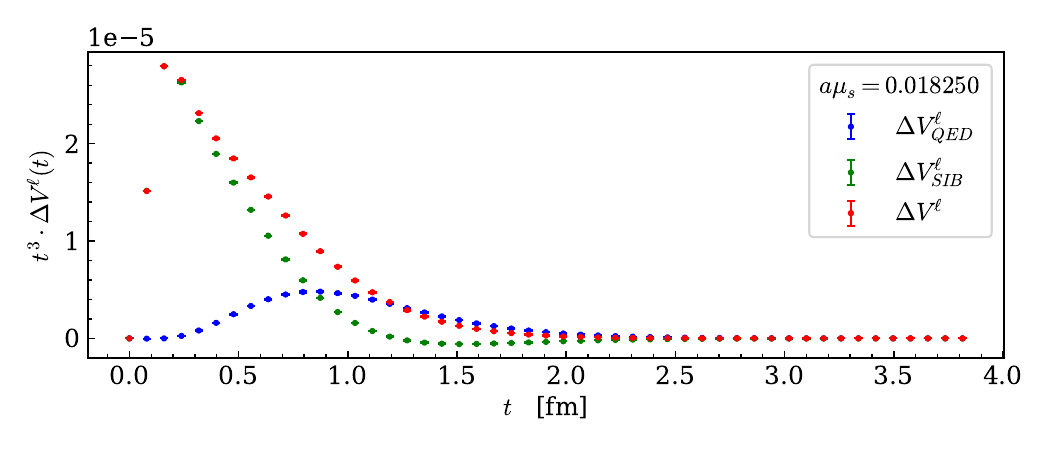}
\includegraphics[width=0.49\linewidth]{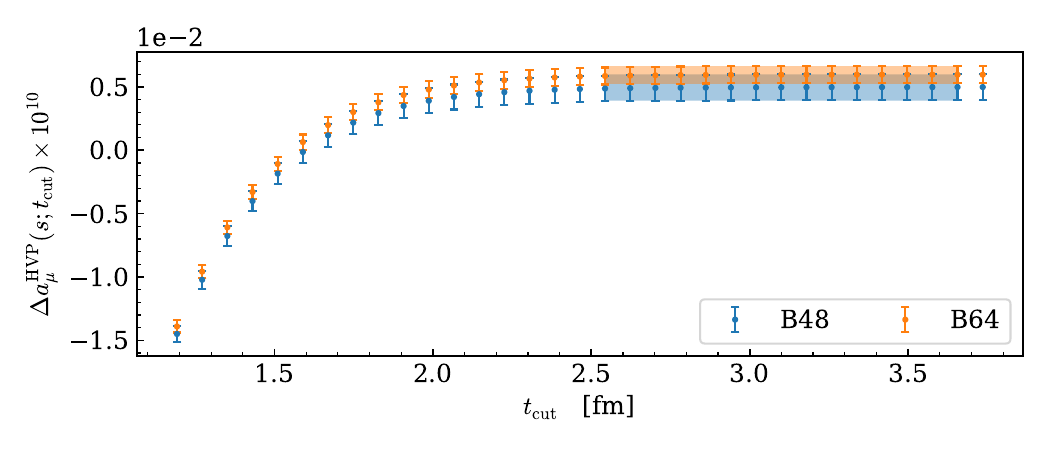} 
\\
\includegraphics[width=0.49\linewidth]{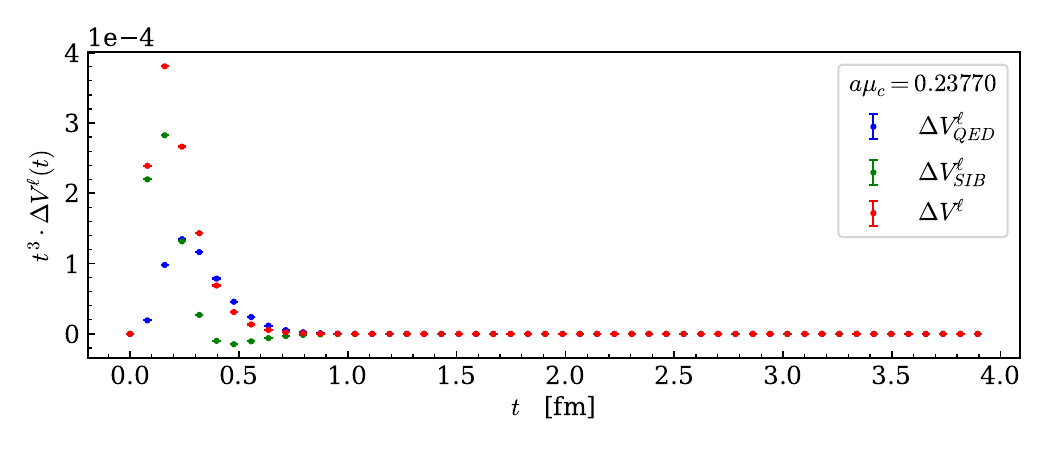}
\includegraphics[width=0.49\linewidth]{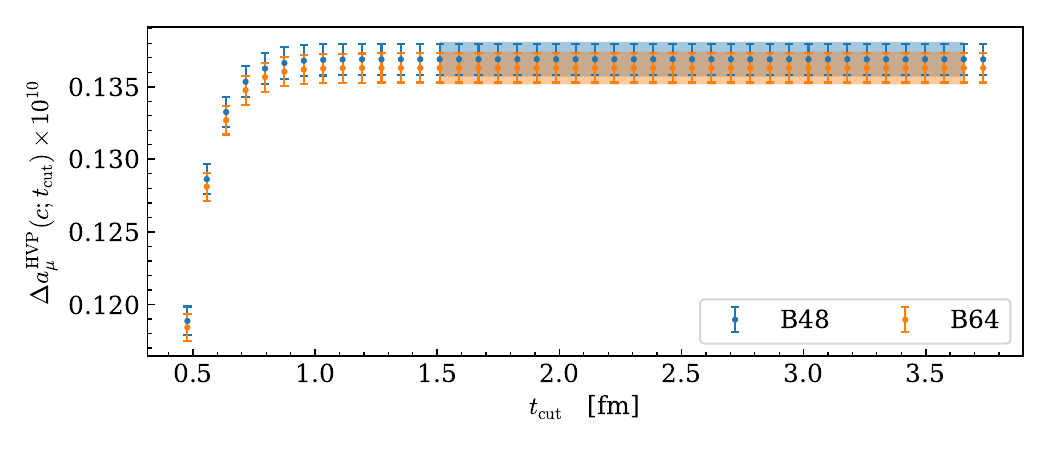} 
\caption{In the left panel, we show the SIB and QED correction to $C_{JJ}^{s}(t)$ and $C_{JJ}^{c}(t)$ for the B64 ensemble. The plateaux analysis of $a_{\mu}^{\mathrm{HVP}}\left(s\right)$ for both the ensembles (red point for B48 and blue points for B64) is shown in the right panel.}
\label{Fig:strange}
\end{figure}

The results for the light contributions are shown in Tab.~\ref{Tab:results}. At the current level of precision, the results on the two volumes are compatible at about two standard deviations.
\begin{figure}[t] 
\centering
\includegraphics[width=0.80\linewidth]{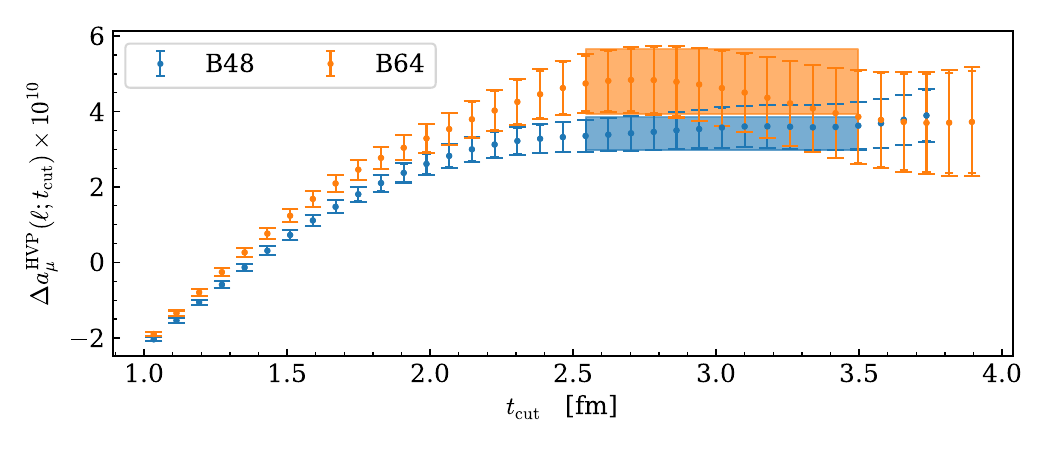}
\caption{The time dependence of the extrapolated $\Delta a_{\mu}^{\mathrm{HVP}}(\ell; t_{\mathrm{cut}})$ on the two volumes $L\sim 3.8$~fm (B48 in blue) and $L\sim 5.1$~fm (B64 in orange).}
\label{Fig:light_extr}
\end{figure}
In the case of strange and charm vector channel correlators, which decrease very quickly with increasing time separation, the degradation of the signal-to-noise ratio is immaterial for the uncertainty of our results (see the left top and bottom panel of Fig.~\ref{Fig:strange}). Indeed, $\Delta a_{\mu}^{\mathrm{HVP}}(s; t_{\mathrm{cut}})$ and $\Delta a_{\mu}^{\mathrm{HVP}}(c; t_{\mathrm{cut}})$ show within errors no dependence on $t_{cut}$ for $t_{\mathrm{cut}} > 2$~fm  and $t_{\mathrm{cut}} > 1$~fm, respecively. We, therefore perform correlated fits to a constant for $t_{\mathrm{cut}}$ values in the range $[2.5, 3.7]$~fm and in the range $[1.5, 3.7]$~fm for the strange and charm quark case, respectively. 

\begin{table}[t]
\centering
\begin{tabular}{rccc}
\toprule
    & $\Delta a_{\mu}^{\mathrm{HVP}}\left(\ell\right) \times 10^{10}$ 
    & $\Delta a_{\mu}^{\mathrm{HVP}}\left(s\right) \times 10^{10}$
    & $\Delta a_{\mu}^{\mathrm{HVP}}\left(c\right) \times 10^{10}$ \\
\midrule
B48    & $3.41 \pm 0.44$ & $0.0049 \pm 0.0010$ & $0.1369 \pm 0.0012$ \\
B64    & $4.79 \pm 0.86$ & $0.0059 \pm 0.0007$ & $0.1363 \pm 0.0011$ \\
\bottomrule
\end{tabular}
\caption{The results of the LIB corrections to $a_{\mu}^{\mathrm{HVP}}\left(\ell\right)$, $a_{\mu}^{\mathrm{HVP}}\left(s\right)$ and $a_{\mu}^{\mathrm{HVP}}\left(c\right)$ contributions for the two different volumes. }
\label{Tab:results}
\end{table}
\section{Conclusions and Outlooks}
In this feasibility study, by using the RM123 approach and working in the electro-quenched approximation, we have computed the LIB corrections to the light, strange and charm quark--connected contributions to the anomalous magnetic moment of the muon. We employed vector-vector two-point correlation functions computed on two ensembles (B48 and B64) with equal lattice spacing ($a\sim 0.08$ fm) and two different linear sizes $L\sim 3.8$~fm, $L\sim5.1$~fm respectively.

The very preliminary results for $\Delta a_{\mu}^{\mathrm{HVP}}\left(\ell\right)$, $\Delta a_{\mu}^{\mathrm{HVP}}\left(s\right)$ and $\Delta a_{\mu}^{\mathrm{HVP}}\left(c\right)$ we present here exhibit an accuracy which is in line with that achieved by other collaborations \cite{Borsanyi:2020mff, Djukanovic:2024cmq, FermilabLattice:2024yho}, as well as ETMC2019 \cite{Giusti:2019xct}. However, our errors here are preliminary and merely statistical and a careful assessment of systematic errors, including possible FSE, is in progress and will be presented in future work. In this context, we plan to repeat this calculation on a third ensemble with equal lattice spacing and a larger volume, corresponding to $L\sim7.6$~fm. A direct comparison of our results with those by other groups is not straightforward since the definition of isoQCD is slightly different and is also deferred to future work. The evaluation of the electro-quenched quark disconnected and electro-unquenched contributions is planned too at three lattice spacings, including the one considered here.

\section*{Acknowlogments}
We thank all members of ETMC for the most enjoyable collaboration. 
S.B. received financial support from the Inno4scale project, which received funding from the European High-Performance Computing Joint Undertaking (JU) under Grant Agreement No. 101118139. The JU receives support from the European Union’s Horizon Europe Programme. The JU receives support from the European Union’s Horizon Europe Programme.
A.E., R.F., G.G, F.S. and N.T. are supported by the Italian Ministry of University and Research (MUR) under the grant PNRR-M4C2-I1.1-PRIN 2022-PE2 Non-perturbative aspects of fundamental interactions, in the Standard Model and beyond F53D23001480006 funded by E.U.- NextGenerationEU. 
M.G. is supported by the Deutsche Forschungsgemeinschaft (DFG, German Research Foundation) as part of the CRC 1639 NuMeriQS – project no. 511713970.
N.K. and R.S. acknowledge support by the Swiss National Science Foundation (SNSF) project No. 200020 208222.
F.S. is supported by ICSC – Centro Nazionale di Ricerca in High Performance Computing, Big Data and Quantum Computing, funded by European Union -NextGenerationEU and by Italian Ministry of University and Research (MUR) projects FIS 00001556 and PRIN 2022N4W8WR.
The authors gratefully acknowledge the Gauss Centre for Supercomputing e.V. (\url{www.gauss-centre.eu}) for funding this project by providing computing time on the GCS Supercomputer JUWELS \cite{JUWELS, JUWELS-BOOSTER} at Jülich Supercomputing Centre (JSC).
We are grateful to CINECA and EuroHPC JU for awarding this project access to Leonardo supercomputing facilities hosted at CINECA. We gratefully acknowledge EuroHPC JU for the computer time on Leonardo-Booster provided to us through the Extreme Scale Access Call grant EHPC-EXT-2024E01-027. We gratefully acknowledge CINECA for the provision of GPU time under the specific initiative INFN-LQCD123.

\bibliographystyle{JHEP}
\bibliography{biblio.bib}

\end{document}